\documentstyle[preprint,aps]{revtex}
\begin{document}
\preprint{LA-UR-93-4317}
\hsize = 7.0in
\widetext
\def\eq{\,=\,}
\def\beq{\begin{equation}}
\def\eeq{\end{equation}}
\def\l{\left(}
\def\r{\right)}
\def\po{\mathaccent 23 p}
\def\pp{p^+}
\def\pr{p_r}
\def\pl{p_\ell}
\draft
\title{
Parity Transformation in the Front Form
\footnotemark[1]
\footnotetext[1]
{The work of D. V. Ahluwalia  was done under the auspices of the U. S.
Department of Energy. He acknowledges the kind support from the Los Alamos
National Laboratory via a postdoctoral research fellowship.
E-mail address for DVA: AV@LAMPF.LANL.GOV}}
\author{Mikolaj Sawicki \ddag \ and D. V. Ahluwalia \dag }

\address{ \ddag Department of Physical Science,
John A. Logan College, Carterville,
Illinois 62918, USA}

\address {\dag Nuclear and Particle Physics Research Group, P-11,
  MS H-846\\ Los Alamos National Laboratory,
 Los Alamos, New Mexico 87545, USA.}

\maketitle
\begin{abstract}
By considering the parity-transformation properties of the $(1/2,\,0)$ and
$(0,\,1/2)$ fields in the {\it front form} we find  ourselves
forced to study the front-form evolution both along $x^+$ and $x^-$ directions.
As a by product, we find that half of the dynamical
degrees of freedom of a full theory live on the $x^+=0$
surface and the other half on the $x^-=0$ surface. Elsewhere, Jacob shows how
these results are required to build a satisfactory, and internally consistent,
front-form quantum field theory.
\end{abstract}

\newpage
In a recent series of papers  \cite{Ra,Rb,Rc,Rd,Re,Rf,Rg,Rh,Ri,Rj},
the
$(j,0)\oplus(0,j)$ representation space has been investigated
in some detail and  some
unexpected results have been discovered. For instance, in Ref. \cite{Ra} it was
found that the $(1,0)\oplus(0,1)$ representation space supports a
Bargmann-Wightman-Wigner-type quantum field theory in which a boson and its
antiparticle have {\it opposite} relative intrinsic parity. In Ref. \cite{Rd},
we applied the approach used previously for the instant-form formalism
\cite{Rc} to the front-form case and obtained the $(j,0)\oplus(0,j)$ spinors
and generalized Melosh transformations for any spin.
The front-form formalism was seen to be endowed with several advantages.  The
work of Ref. \cite{Rd}, apart from the indicated generalization, reproduced
some of the well-known results of Melosh \cite{M}, Lepage and Brodsky
\cite{LB}, and Dziembowski \cite{D} for spin-$1\over2$. Due to certain magic of
Wigner's time-reversal operator in Ref. \cite{Ri}, we were able to present a
Majorana-like construct in the $(j,0)\oplus(0,j)$ representation space. Even
though much further work  is needed to exploit the Majorana-like construct,  we
suspect that for spin-$1$ it may have a deep connection with
$\cal P$ and $\cal CP$
violation \cite{Rj} in nature.

During the  above-indicated investigations,
we have come across a problem in the
front form of field theory. The problem deals with the operation of parity. It
is the purpose of this paper to bring
attention to this problem and present its
solution. We note that it is not  the first time that the problem of parity
in the front form has been considered. What we present is  a new and
significantly more critical analysis of the subject. For example, one of the
earliest considerations of this problem appears in the 1971 thesis of Soper
\cite{Soper}. More recently, Jacob \cite{OJ} has considered the question of
parity while considering the quantization of
the scalar field, and he arrived at similar conclusions to those
presented in this
paper {\it but} with a very different perspective. We look at the
transformation properties of the $(1/2,\,0)$ and $(0,\,1/2)$ fields in the
front form and find that the consideration of parity-covariance {\it forces} us
to consider evolution not only along $x^+$ (or $x^-$) but simultaneously along
$x^+$ and $x^-$. Our considerations are applicable to massive as well as
massless particles. McCartor \cite{GM}, while investigating the quantization of
massless fields in the front form concluded that a spin-$1\over 2$ system must
be specified on  both $x^+$ and $x^-$ surfaces. Later, McCartor and Robertson
\cite{MR} argued that $x^+$ and  $x^-$ should be considered symmetrically. The
arguments that we construct below lie at the heart of space-time symmetries
(and are independent of any specific Lagrangian), and the reader should find
that these arguments convincingly establish that parity-covariance {\it
requires} studying the evolution of a system (in the front form of field
theory) both along $x^+$ and $x^-$ directions. The arguments we present are
true for {\it all} $(j,0)$ and $(0,j)$ fields. However, for conceptual clarity
and general familiarity, we choose spin-$1\over 2$ as an example case.

To define the problem,
we recall  that in the {\it instant form} $({1/2},\,0)$ and
$(0,\,{1/ 2})$ fields transform as \cite{Ra,Rc,LR}
\begin{eqnarray} &&\left({1/
2},\,0\right):\quad \phi_{_R}(p^\mu)\,=\, \exp\left[+\, {\bbox \varphi}\cdot
{{\bbox \sigma}\over 2} \right]\,\phi_{_R}({\overcirc p}^\mu)\quad,\nonumber\\
&&\left(0,\,{1/ 2}\right):\quad \phi_{_L} (p^\mu)\,=\,\exp\left[ -\, {\bbox
\varphi}\cdot{{\bbox \sigma}\over 2} \right]\,\phi_{_L}({\overcirc p}^\mu)
\quad. \label{if} \end{eqnarray}
In the above equations, $p^\mu$ represents the
four-momentum of the particle and ${\overcirc p}^\mu$ corresponds to the
particle at rest. The boost operator $\bbox \varphi$ that appears in Eq.
(\ref{if}) is defined
 as
\begin{equation} \cosh(\varphi\,)
\,=\,\gamma\,=\,{1\over\sqrt{1-v^2}}\,=\,{E\over m},\quad \sinh(
\varphi\,)\,=\,v\gamma\,=\,{| {\bf p}\, |\over m},\quad \hat{\bbox  \varphi}
\,=\,{ {\bf p} \over {|{ \bf p}\,|}},\quad
\varphi\,=\,\vert{\bbox\varphi}\vert\quad, \label{bp} \end{equation} with $\bf
p$ the three-momentum of the particle. It is immediately obvious from Eqs.
(\ref{if}) and (\ref{bp}) that under the operation parity, $\cal P$,
$({1/2},\,0)$ and $(0,\,{1/2})$ representation spaces  get interchanged,
\begin{equation} {\cal P}:\quad \left({1/2},\,0\right)\,\leftrightarrow\,
\left(0,\,{1/ 2}\right)\quad.\label{ifp} \end{equation}
It is because of the
result (\ref{ifp}) that any parity-conserving interaction must involve both the
$(1/2,\,0)$ and $(0,\,1/2)$ fields.
One then  introduces the $(1/2,\,0)\oplus(0,\,1/2)$ Dirac spinor,
which in the chiral representation (indicated by the use of curly brackets
enclosing the argument $p^\mu$) reads
\begin{equation}
\psi\{p^\mu\}\,=\,
\left[
\begin{array}{c}
\phi_{_R}(p^\mu)\\
\phi_{_L}(p^\mu)
\end{array}\right]\quad.
\end{equation}
These results are well known and can indeed be found in any modern textbook
on
quantum field theory \cite{LR,MK,GS}. For the sake of later reference,
let's note that the more familiar \cite{BD} canonical representation
is defined as (argument $p^\mu$ of spinors in the canonical representation
would be enclosed in square brackets)
\begin{equation}
\psi[\,p^\mu]\,=\,
{1\over{\sqrt{2}}}\,\left[
\begin{array}{ccc}
\openone &{\,\,\,}& \openone\\
\openone &{\,\,\,}& -\openone
\end{array}
\right]\,
\psi\{p^\mu\}\quad,\label{s}
\end{equation}
where $\openone$ is a $2\times 2$ identity matrix. Under the operation of
parity operator $S({\cal P})=\gamma^0$ in the $(1/2,\,0)\oplus(0,\,1/2)$
representation space, the particle and antiparticle spinors, in the usual
notation of Refs. \cite{Rc,BD}, transform as
\begin{eqnarray}
u_\sigma[\,p^{\prime\,\mu}] \,=\,+\,\gamma^0\,u_\sigma[\,p^\mu]\quad,\nonumber
\\
v_\sigma[\,p^{\prime\,\mu}]
\,=\,-\,\gamma^0\,v_\sigma[\,p^\mu]\quad.\label{ifuv}
\end{eqnarray}
The $p^{\prime\,\mu}$ is the parity-transformed $p^\mu$.

The front-form
counterpart of the simple and important instant-form relation (\ref{ifp}),
and other equations such as (\ref{ifuv}),  is a little subtle.
To see this note, that counterpart of transformation properties of the right-
and left-handed fields in the front-form of evolution {\it associated with
$x^+=x^0+x^3$} reads (as was recently shown in Ref. \cite{Rd})
\begin{eqnarray}
&&\left({1/ 2},\,0\right)^{[x^+]}:\quad \phi^{[x^+]}_{_R}(p^\mu)\,=\,
\exp\left[+ \,{\bbox\beta}\cdot  {{\bbox \sigma}\over 2}
\right]\,\phi^{[x^+]}_{_R}({\overcirc p}^\mu)\quad,\nonumber\\
&&\left(0,\,{1/ 2}\right)^{[x^+]}:\quad \phi^{[x^+]}_{_L}
(p^\mu)\,=\,\exp\left[ - \,{\bbox\beta^\ast}
\cdot{{\bbox \sigma}\over 2}
\right]\,\phi^{[x^+]}_{_L}({\overcirc p}^\mu) \quad. \label{ff}
\end{eqnarray}
The superscript $[x^+]$ in the above equations serves the purpose of reminding
that these relations hold true for the evolution along $x^+$. The
${\bbox\sigma}$ are the standard Pauli matrices. The boost parameter
${\protect\bbox \beta}$ that appears in Eq. (\ref{ff})
is defined \cite{Rd}  as:
\begin{equation}
\protect\bbox{ \beta}
\,=\, \eta\,\left(\alpha\,v^r\,,\,\,-i\,\alpha\,  v^r\,,\,\,1\right)\quad,
\end{equation}
where
$
\alpha\,=\,\left[1\,-\,\exp(-\eta)\right]^{-1}\,,
$
 $v^r\,\,=\,\,v_x\,+\,i\,v_y$ (and
$v^\ell\,\,=\,\,v_x\,-\,i\,v_y$). In terms of the front-form variable
$p^+ \equiv E+p_z$, one can show that
\begin{equation}
\cosh(\eta/2)=\Omega
\left(p^+ + m\right)\,,\,
\sinh(\eta/2)=\Omega
\left(p^+ - m\right)\,,
\end{equation}
with
$
\Omega\,=\,\left[1/ (2 m)\right]\sqrt{m/ {p^+}}\,.
$
Under the operation of parity, $\cal P$,  an  inspection of Eqs. (\ref{ff}),
indicates that
$({1/2},\,0)$ and $(0,\,{1/2})$ representation
spaces do {\it not}  get interchanged;
\begin{equation}
{\cal P}:
\quad \left({1/2},\,0\right)\,\not\leftrightarrow\, \left(0,\,{1/
2}\right)\quad, \quad {\mbox{\rm for evolution along $x^+\,=\,x^0+x^3$}}
\quad.\label{ffp}
\end{equation}
Therefore, it is obvious that the standard front-form formulation (which
expands
field operators in terms of the spinors associated  only with the evolution
along the $x^+\,=\,x^0+x^3$, {\it or } $x^-\,=\,x^0-x^3$ directions) of field
theories is not covariant under parity.  Added to this observation is the fact
that only {\it half} of the degrees of freedom associated
with such a field are dynamical
\cite{CRY}. We claim that there is in fact no loss/reduction  of degrees of
freedom in the front form of field theories and no violation of parity
covariance {\it if} one is a little careful. We now elaborate.

One must note that under the operation of parity, $\cal P$, unlike  the
instant-form case, the direction of evolution (say) $x^+$ gets interchanged
with $x^-$. So, we  suspect that one must
obtain counterparts of transformation properties (\ref{ff}) for the evolution
along  the parity-transformed $x^+$, that is $x^-$, and
see how the fields transform.
Algebraically, this exercise is
 a little involved but follows parallel to our previous analysis of
Ref. \cite{Rd}.
Here, we just quote the result of our calculations.
We find that the right- and left-handed fields in the
front form of evolution {\it associated with $x^-$} direction transform as
follows
\begin{eqnarray}
&&\left({1/ 2},\,0\right)^{[x^-]}:\quad \phi^{[x^-]}_{_R}(p^{\prime\mu})\,=\,
\exp\left[-\, {\bbox\beta^\ast}\cdot  {{\bbox \sigma}\over 2}
\right]\,\phi^{[x^-]}_{_R}({\overcirc p}^\mu)\quad,\nonumber\\ &&\left(0,\,{1/
2}\right)^{[x^-]}:\quad \phi^{[x^-]}_{_L} (p^{\prime\mu})\,=\,\exp\left[ +\,
{\bbox\beta} \cdot{{\bbox \sigma}\over 2} \right]\,\phi^{[x^-]}_{_L}({\overcirc
p}^\mu) \quad. \label{ffb}
\end{eqnarray}
The superscript $[x^-]$ in the above
equations serves the purpose
of reminding that these relations hold true for the evolution along $x^-$.

Comparison of transformation properties (\ref{ff}) and (\ref{ffb}) yields the
front-form counterpart of the instant-form relation (\ref{ifp}),
\begin{equation}
{\cal P}:\quad
\left\{
\begin{array}{l}
(1/2,\,0)^{[x^+]}\,\leftrightarrow \,(0,\,1/2)^{[x^-]}\\
(0,\,1/2)^{[x^+]}\,\leftrightarrow \,(1/2,\,0)^{[x^-]}
\end{array}
\right.\quad.
\end{equation}
Thus, under the operation of parity, the representation space
$(1/2,\,0)^{[x^+]}\oplus (0,\,1/2)^{[x^+]}$ maps one-to-one onto
$(1/2,\,0)^{[x^-]}\oplus (0,\,1/2)^{[x^-]}$. To be more explicit, one may carry
out an exercise similar to the one presented in our recent work
\cite{Rd} and obtain the
$u^{[x^-]}_{\mu}[\,p^{\prime\,\mu}]$ and
$v^{[x^-]}_{\mu}[\,p^{\prime\,\mu}]$
spinors in the $(1/2,\,0)^{[x^-]}\oplus
(0,\,1/2)^{[x^-]}$ representation space. We already know the
$(1/2,\,0)^{[x^+]}\oplus (0,\,1/2)^{[x^+]}$-spinors,
$u^{[x^+]}_h[\,p^{\prime\,\mu}]$ and
$v^{[x^+]}_h[\,p^{\prime\,\mu}]$, from Ref. \cite{Rd}. The
parity operation $S({\cal P})$ still remains $\gamma^0$ because Melosh
transformation, as was  explicitly proved in \cite{Rd}, does not mix
particle and antiparticle spinors, {\it and} the front-form
$(1/2,\,0)\oplus(0,\,1/2)$ spinors turn out to be the superposition of the
instant-form spinors with $p^\mu$-dependent coefficients contained in the
Melosh matrix for spin-$1\over 2$. The above indicated exercise yields the
front-form counterpart of the identities (\ref{ifuv}),
\begin{eqnarray}
u^{[x^-]}_{\mu}[\,p^{\prime\,\mu}]
\,=\,+\,\gamma^0\,u^{[x^+]}_h[\,p^\mu]\quad,\nonumber \\
v^{[x^-]}_{\mu} [\,p^{\prime\,\mu}]
\,=\,-\,\gamma^0\,v^{[x^+]}_h[\,p^\mu]\quad.\label{ffuv}
\end{eqnarray}
The $h $ and $\mu$ in the above expressions correspond to the helicity
degrees of freedom associated with the front-form helicity operators
(respectively associated with evolution along $x^+$ and $x^-$):
\begin{eqnarray}
{\cal J}^{[x^+]}_3 \equiv J_3 \,+\,{1\over
P_-}\left(G_1\,P_2\,-\,G_2\,P_1\right)\quad,
\\
{\cal J}^{[x^-]}_3 \equiv J_3 \,+\,{1\over P_+}\left(D_1\,P_2\,-\,D_2\,
P_1\right)\quad.
\end{eqnarray}
For dynamical significance and  the definition of various generators involved
in the above expressions, we refer
 the reader  to Sec. II of Ref. \cite{Rd}.
The  non-trivial space-time structure of the results we obtain, such as
Eqs. (\ref{ffuv}), is  intuitively satisfactory. The reader should carefully
examine all the associated super- and sub- scripts !

To conclude, we note that evolution along the $x^+$  direction contains
two dynamically independent spinorial degrees of freedom \cite{CRY}.
It can now be verified, following arguments similar to \cite{CRY}, that the
other two dynamically independent degrees of freedom are contained in the field
operator constructed for the evolution along $x^-$. A parity-covariant
front-form field theory requires specification of a system on both $x^+=0$ and
$x^-=0$ surfaces and carries {\it four} independent degrees of freedom (as in
the instant-form) --- {\it two} on each surface.  Jacob \cite{OJb}, in
a recent preprint, shows that the result (\ref{ffuv}) is essential for
front-form quantization, which involves specification of a system on both
$x^+=0$ and $x^-=0$ surfaces.

\acknowledgements
It is our pleasure to thank Ovid  Jacob  (SLAC) for insightful conversations on
the general subject of this work.

\end{document}